\documentclass[a4paper,11pt]{article}
\pdfoutput=1 

\usepackage{jcappub} 

\usepackage[T1]{fontenc} 
\usepackage[utf8]{inputenc}

\usepackage{diagbox}
\usepackage{color}
\usepackage{multirow}
\usepackage{hhline}

\usepackage{caption}
\usepackage{subcaption}

\title{Space-borne atom interferometric gravitational wave detections. Part I. The forecast of bright sirens on cosmology}

\author[a,b,c]{Rong-Gen Cai}
\emailAdd{cairg@itp.ac.cn}
\author[d,1]{Tao Yang %
\note{Corresponding author}}
\emailAdd{yangtao.lighink@gmail.com}

\affiliation[a]{CAS Key Laboratory of Theoretical Physics, Institute of Theoretical Physics, Chinese Academy of Sciences, Beijing 100190, China}
\affiliation[b]{School of Physical Sciences, University of Chinese Academy of Sciences, No.19A Yuquan Road, Beijing 100049, China}
\affiliation[c]{School of Fundamental Physics and Mathematical Sciences, Hangzhou Institute for Advanced Study (HIAS), University of Chinese Academy of Sciences, Hangzhou 310024, China}

\affiliation[d]{Asia Pacific Center for Theoretical Physics, Pohang 37673, Korea}

\abstract{Atom interferometers (AIs) as gravitational-wave (GW) detectors have been proposed a decade ago. Both ground and space-based projects will be in construction and preparation in the near future. In this paper, for the first time, we investigate the potential of the space-borne AIs on detecting GW standard sirens and hence the applications on cosmology. We consider AEDGE as our fiducial AI GW detector and estimate the number of bright sirens that would be obtained within a 5-years data-taking period of GW and with the follow-up observation of electromagnetic (EM) counterparts. We then construct the mock catalogue of bright sirens and predict their ability on constraining cosmological parameters such as the Hubble constant, dynamics of dark energy, and modified gravity theory. Our preliminary results show around order $\mathcal{O} (30)$ bright sirens can be obtained from a 5-years operation time of AEDGE and the follow-up observation of EM counterparts. The bright sirens alone can measure $H_0$ with a precision of 2.1\%, which is sufficient to arbitrate the Hubble tension. Combining current most precise electromagnetic experiments, the inclusion of AEDGE bright sirens can improve the measurement of the equation of state of dark energy, though marginally. Moreover, by modifying GW propagation on cosmological scales, the deviations from general relativity (modified gravity theory effects) can be constrained at 5.7\% precision level. }


\begin{document}
\maketitle
\flushbottom

\section{Introduction}
The laser interferometer (LI) experiment of gravitational waves (GWs) conducted by LIGO/Virgo has achieved great success in the past five years. Up to now, LIGO has reported more than 50 confirmed GW events produced by the merger of the binary black holes (BBH), of the binary neutron stars (BNS), and of the neutron star-black hole binary (NS-BH)~\cite{LIGOScientific:2016aoc,LIGOScientific:2017vwq,LIGOScientific:2018mvr,LIGOScientific:2020ibl,LIGOScientific:2021qlt}. Especially, the first joint observation of GW from a BNS with its electromagnetic counterpart (EM) has opened the new era of multimessenger astronomy~\cite{LIGOScientific:2017vwq,LIGOScientific:2017ync,LIGOScientific:2017zic}. GWs as the novel signals play significant roles in  modern cosmology, astrophysics, and fundamental physics (see reviews~\cite{Schutz:1999xj,Barack:2018yly,Sasaki:2018dmp,Gair:2012nm,Ezquiaga:2018btd,Cai:2017cbj,Meszaros:2019xej,Christensen:2018iqi,Perkins:2020tra}, and a very recent review of the progress in GW physics~\cite{Bian:2021ini}). In the next few years, the second generation ground-based GW detector network HLVKI (consisting of advanced LIGO-Hanford, advanced LIGO-Livingston, advanced Virgo, KAGRA and LIGO-India) will be in operation. On a longer timescale, around the 2030s, the third-generation ground-based detectors such as  Einstein Telescope (ET)~\footnote{\url{http://www.et-gw.eu/}} and Cosmic Explorer (CE)~\footnote{\url{https://cosmicexplorer.org/}}, and the space interferometer LISA~\footnote{\url{https://www.lisamission.org/}} will start to work. During the same time, Chinese space-based GW detectors like Taiji~\cite{Hu:2017mde,Ruan:2018tsw} and Tianqin~\cite{TianQin:2015yph} would also be launched. We expect a joint network of LI GW experiments from ground to space in the coming two decades. 

For the ground-based detectors, typically their sensitive frequencies are $f>10$ Hz, while for space-based experiments $f<0.1$ Hz. Atom interferometers (AIs) are the candidates to probe GW in the Deci-Hz gap between LIGO/Virgo and LISA. The concept of AI as GW detector has been proposed a decade ago~\cite{Dimopoulos:2007cj,Dimopoulos:2008sv,Graham:2012sy,Hogan:2015xla}. The AI projects such as ground based ZAIGA~\cite{Zhan:2019quq} in China, AION~\cite{Badurina:2019hst} in  UK, MIGA~\cite{Geiger:2015tma} in France, ELGAR~\cite{Canuel:2019abg} in Europe, and the space-borne MAGIS~\cite{Graham:2017pmn} and AEDGE~\cite{AEDGE:2019nxb} have been proposed and in preparation. 

AIs can be used as probes for both dark matter and gravitational waves (see the introductions in~\cite{Graham:2017pmn,Badurina:2019hst,AEDGE:2019nxb}). Several specific research of AIs on such as the localization of GW sources~\cite{Graham:2017lmg} and the constraints of the deviations from general relativity (GR)~\cite{Ellis:2020lxl} have been investigated. The frequency window of AIs between 0.01 and a few Hz is ideal for the observation of mergers involving intermediate-mass black holes (IMBHs) with masses in the range 100 to $10^5$ solar masses. However it can also observe the early stage of the inspiral phase of BBH and BNS (see examples in~\cite{Graham:2017lmg,Ellis:2020lxl}).  Take the space-borne AEDGE as an example, the motion of the detectors around the Sun as well as in Earth orbit would provide a very precise angular localization thus providing early warning of possible upcoming multimessenger events. There could be significant synergies between AEDGE measurements and observations in other frequency ranges by such as LISA, LIGO and ET.

In this paper, to fill the gap in cosmological applications of AIs, we would like to investigate the potential of the standard sirens by AIs as probes of cosmology and modified gravity. We focus on the bright sirens, i.e., BNS associated with EM counterparts. The coincidence between GW and EM observations is the most straightforward way to obtain both the luminosity distance and redshift information thus constructing the $d_L-z$ relation for constraining cosmological parameters~\cite{Dalal:2006qt,Nissanke:2009kt,Zhao:2010sz,Cai:2016sby,Cai:2017yww,Yang:2021qge}. We use the space-borne AEDGE as the fiducial AI GW detector.  The basic design of AEDGE requires two satellites operating along a single line-of-sight and separated by a long distance. The payload of each satellite will consist of cold atom technology as developed for state-of-the-art atom interferometry and atomic clocks~\cite{AEDGE:2019nxb}. This project assumes a minimum data-taking time of 3 years, which requires a mission duration of at least 5 years, while 10 years would be an ultimate goal.
As a first step, we would like to estimate how many BNS and joint GW+EM events can be obtained by AEDGE and the follow-up EM observations in the future. With these events, how precisely one can measure the cosmological parameters such as the Hubble constant and constrain the feature of dark energy and deviations from GR, which are the key issues we would like to investigate in this paper.

The structure of this paper is as follows. In section~\ref{sec:const_StS} we construct the catalogue of the bright sirens by AEDGE in 5-years operation time and with the follow-up observation of EM counterpart. Then the Hubble diagram is built accordingly. We apply the mock bright sirens to several applications in cosmology in section~\ref{sec:cosmology}. We are pretty interested in the measurement of Hubble constant, the constraints of the dynamics of dark energy, and the deviations from GR (modified gravity effect). Finally, we draw the conclusions and give some discussions in~\ref{sec:conclusion}.

\section{Construction of BNS bright sirens \label{sec:const_StS}}
Following~\cite{Vitale:2018yhm,Belgacem:2019tbw,Yang:2021qge}, we adopt the merger rate of massive binaries $R_m$ by assuming an exponential time delay distribution $P(t_d,\tau)=\frac{1}{\tau}\exp(-t_d/\tau)$ with an e-fold time of $\tau=100$ Myr~\cite{Vitale:2018yhm},
\begin{equation}
R_m(z_m)=\int_{z_m}^{\infty}dz_f\frac{dt_f}{dz_f}R_f(z_f)P(t_d) \,.
\label{eq:Rm}
\end{equation}
Here $t_m$ (or the corresponding redshift $z_m$) and $t_f$ are the look-back time when the systems merged and formed. $t_d=t_f-t_m$ is the time delay. $R_f$ is the formation rate of massive binaries and we assume it is proportional to the Madau-Dickinson (MD) star formation rate~\cite{Madau:2014bja},
\begin{equation}
\psi_{\rm MD}=\psi_0\frac{(1+z)^{\alpha}}{1+[(1+z)/C]^{\beta}} \,,
\label{eq:psiMD}
\end{equation}
with parameters $\alpha=2.7$, $\beta=5.6$ and $C=2.9$. The coefficient $\psi_0$ is the normalization factor which is determined by the BNS rate we set at $z=0$. We adopt $R_m(z=0)=920~\rm{Gpc}^{-3}\rm{yr}^{-1}$ which is the median rate estimated from the O1 and O2 LIGO/Virgo observation run~\cite{LIGOScientific:2018mvr} and assume a Gaussian distribution of the mass of neutron stars. It is also consistent with the first half O3 run~\cite{LIGOScientific:2020ibl}. Then we can convert the merger rate per volume in the source frame to merger rate density per unit redshift in the observer frame
\begin{equation}
R_z(z)=\frac{R_m(z)}{1+z}\frac{dV(z)}{dz} \,,
\label{eq:Rz}
\end{equation}
where $dV/dz$ is the comoving volume element.

Having the BNS merger rate $R_z(z)$ in Eq.~(\ref{eq:Rz}), we can first sample the BNS mergers from redshift 0 to 0.5 in a Monte-Carlo way. For every merger, we assign the sky location ($\theta$, $\phi$) and inclination angle $\iota$ from isotropic distribution.  The polarization $\psi$, component masses of BNS are drawn from the uniform distribution, i.e., $\psi\in[0,2\pi)$ and $m_1,m_2\in[1,2]M_{\odot}$. Then we can calculate the signal-to-noise ratio (SNR) in the inspiral phase
\begin{equation}
\rho^2=\frac{5}{6}\frac{(G\mathcal{M}_c)^{5/3}\mathcal{F}^2}{c^3\pi^{4/3}d_L^2(z)}\int^{f_{\rm max}}_{f_{\rm min}}df\frac{f^{-7/3}}{S_n(f)} \,,
\label{eq:SNR}
\end{equation}
where $\mathcal{M}_c=(m_1m_2)^{3/5}/(m_1+m_2)^{1/5}(1+z)$ is the redshifted chirp mass. $d_L$ is the luminosity distance. $S_n(f)$ is the one-sided noise power spectral density (PSD) of detector. The factor $\mathcal{F}$ is to characterize the detector response, $\mathcal{F}^2=\frac{(1+\cos^2\iota)^2}{4}F^2_++\cos^2\iota F^2_\times$. $F_+$ and $F_\times$ are the antenna response functions to the  $+$ and $\times$ polarizations of GW. For lower and upper limits of the frequency in the integral of Eq.~(\ref{eq:SNR}),  we adopt the frequency window of AEDGE (0.01-3 Hz) in the resonant modes~\cite{AEDGE:2019nxb}. This frequency range is far from the merger phase of BNS so we do not need to set a cut-off to make sure the post-Newtonian approximation of the waveform is always valid. In Eq.~(\ref{eq:SNR}), we adopt the noise PSD of AEDGE in the resonant modes.
There are two modes for the operation of AIs like MAGIS~\cite{Graham:2017pmn} and AEDGE~\cite{AEDGE:2019nxb}. For the resonant mode, the frequency band is very narrow at a specific value, the observing strategy is to first sit at the lower limit of the frequency window of AEDGE, waiting for a source to enter the band. Once a source is discovered it can be tracked for longer by sweeping the detector frequency up to follow the source. The optimal strategy for tracking hundreds of BNS using the resonant modes is still under debate and out of the scope of this paper. An alternative way is using the broadband modes like the traditional LIs. The broadband modes are usually less sensitive than the resonant modes. But it is not clear how  inferior the sensitivity will be, comparing to the case of resonant modes of AEDGE when it is in operation. However, in this paper, we adopt the sensitivity curve of AEDGE in resonant modes (see the envelope of the curve in figure 1 of~\cite{Ellis:2020lxl}) and we consider this to be an optimistic case. 
The final SNR is accumulated along the tracking of BNS signal and we just integrate it from the lower to upper frequency limits in Eq.~(\ref{eq:SNR})\footnote{Actually we do not need to integrate Eq.~(\ref{eq:SNR}) in the full range of the frequency limits. The SNR is accumulated very quickly in the range from 0.08 to 3 Hz (the most sensitive region,  dominating the total value of SNR). So one can flexibly choose the starting point of the integral (around 0.08 Hz) for the calculation of SNR. Thus the average time for tracking one BNS would be much less. It will benefit the time allocation of observing many BNS when using resonant modes. }.

To calculate the explicit form of the antenna response functions, we follow~\cite{Graham:2017lmg} and consider the single-baseline detector AEDGE measuring the gravitational stretching and contraction along its baseline direction. Thus the detector response tensor $D_{ij}$ from the baseline direction unit vector $a_i(t)$ is
\begin{equation}
D_{ij}=\frac{1}{2}a_i(t)a_j(t) \,.
\label{eq:Dij}
\end{equation}
Then the GW strain tensor can be decomposed in terms of 
\begin{equation}
h_{ij}(t)=h_+(t)e_{ij}^++h_\times(t)e_{ij}^{\times}\,,
\label{eq:hij}
\end{equation}
here $e_{ij}^{+,\times}$ are  the polarization tensors with $e_{ij}^+=\hat{X}_i\hat{X}_j-\hat{Y}_i\hat{Y}_j$, and $e_{ij}^\times=\hat{X}_i\hat{Y}_j+\hat{Y}_i\hat{X}_j$. The basis is~\cite{Nissanke:2009kt,Cutler:1994ys},
\begin{align}
\hat{X}=&(\sin\phi\cos\psi-\sin\psi\cos\phi\cos\theta, -\cos\phi\cos\psi-\sin\psi\sin\phi\cos\theta, \sin\psi\sin\theta) \,,\\
\hat{Y}=&(-\sin\phi\sin\psi-\cos\psi\cos\phi\cos\theta, \cos\phi\sin\psi-\cos\psi\sin\phi\cos\theta, \cos\psi\sin\theta) \,.
\end{align}
Then the observed waveform is given by
\begin{equation}
h(t)\equiv D_{ij}h_{ij}=h_+(t)F_+(t)+h_\times(t)F_\times(t) \,,
\end{equation}
with $F_+(t)=D_{ij}(t)e_{ij}^+$ and $F_\times(t)=D_{ij}(t)e_{ij}^\times$. Since the atom GW detector reorients and/or moves along the orbit around the Earth, the observed waveform and phase are modulated and Doppler-shifted, yielding important angular information. Without loss of generality, we can parameterize the detector location on the orbit by a unit vector $r_0(t)=(\cos\phi_a(t), \sin\phi_a(t), 0)$, where $\phi_a(t)=2\pi t/T_{\rm AI}+\phi_0$ is the azimuthal orbit angle around the Earth. $T_{\rm AI}$ is the AEDGE orbit period. The baseline direction of AEDGE is $a_0(t)=(-\sin\phi_a(t), \cos\phi_a(t), 0)$. Now we should transform the Earth polar coordinate of $r_0(t)$ and $a_0(t)$ to the Sun’s polar coordinate as
\begin{align}
r_{\rm AI}(t)=
\begin{pmatrix}
\cos\phi_{\rm Ea}(t) & -\sin\phi_{Ea}(t) & 0 \\
\sin\phi_{\rm Ea}(t) & \cos\phi_{Ea}(t) & 0 \\
0 & 0 & 1
\end{pmatrix} \cdot
\begin{pmatrix}
\cos\theta_{\rm inc} & 0 & -\sin\theta_{\rm inc} \\
0 & 1 & 0 \\
\sin\theta_{\rm inc} & 0 & \cos\theta_{\rm inc}
\end{pmatrix}
\cdot
r_0(t) \,.
\end{align}
The baseline direction $a(t)$ is transformed similarly. The azimuthal angle of the Earth’s orbit around the Sun is $\phi_{\rm Ea}=2\pi t/(1 \rm yr)+\phi_0'$. $\theta_{\rm inc}$ is the inclination. We adopt a similar setup for AEDGE with MAGIS~\cite{Graham:2017pmn,Ellis:2020lxl}, i.e., the detector consists of two satellites in identical circular geocentric orbits with an inclination $\theta_{\rm inc}=28.5^{\circ}$ at an altitude of $24\times10^3$ km and forming a baseline of $44\times10^3$ km with $T_{\rm AI} = 10$ hours.

\begin{figure}
\centering
\includegraphics[width=0.49\textwidth]{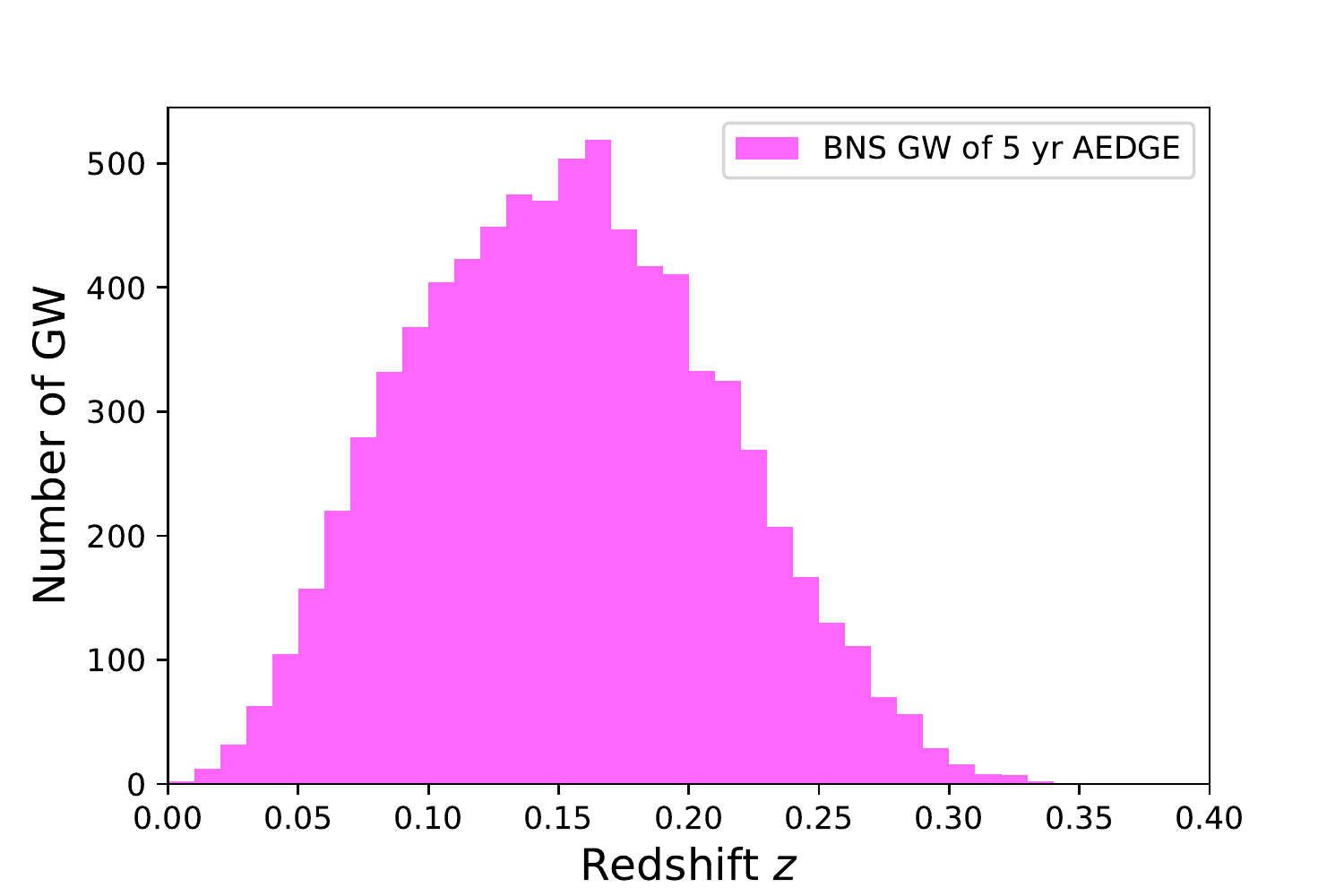}
\includegraphics[width=0.49\textwidth]{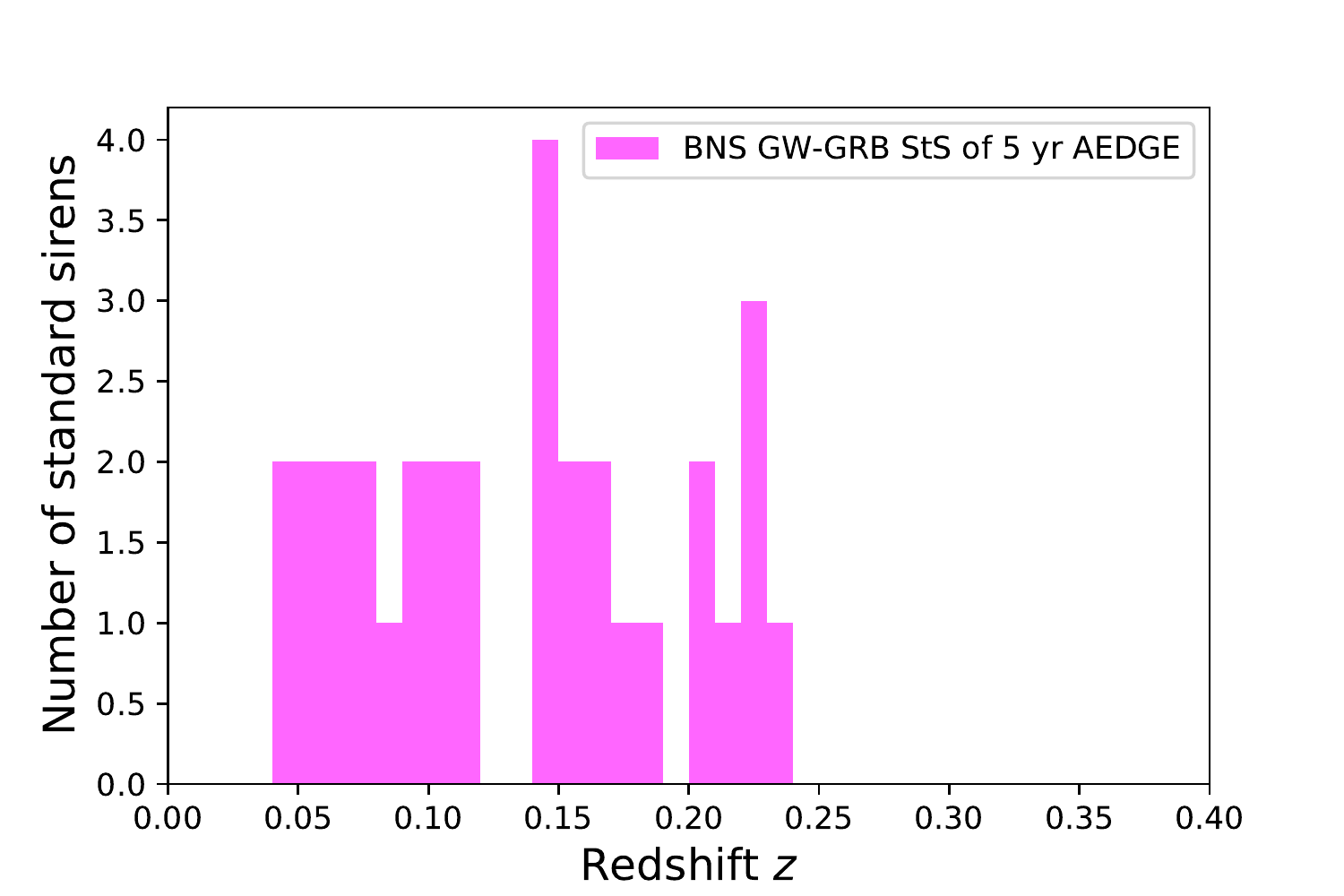}
\caption{A realization of the mock catalogue of 5-years detections of BNS GW (left) and GW-GRB standard sirens  (StS) (right) from AEDGE assuming a GRB detector with the characteristics of THESEUS.}
\label{fig:BNSGW}
\end{figure}

Now the derivation of the antenna response functions is straightforward. For every sampled BNS merger we calculate the SNR from Eq.~(\ref{eq:SNR}) and select the BNS events with a threshold SNR>8. The histogram of the selected BNS events is shown in the left panel of Fig.~\ref{fig:BNSGW}. We can see that  AEDGE can detect about 7819 BNS cases (1564/yr) from redshift 0--0.34 in a 5-years data-taking period, which can be compared to the second-generation ground-based detector network HLVKI with 86/yr from redshift 0--0.15~\cite{Belgacem:2019tbw,Yang:2021qge}, and third-generation network ET+2CE with $\mathcal{O}(10^6)$/yr from redshift 0--9~\cite{Yang:2021qge}.

To use BNS as the bright sirens, we need to select the events with EM counterparts from the total BNS events. 
We adopt the similar strategy as in~\cite{Yang:2021qge} and consider GW+EM counterpart as the coincidence of GW event by AEDGE and later (even years later) follow-up observation of short GRBs by EM telescope. A GRB detected in coincidence with a GW signal requires that the peak flux is above the flux limit of the satellite. Considering the construction time of AEDGE, we select the THESEUS mission~\cite{THESEUS:2017qvx,THESEUS:2017wvz} as the GRB satellite to predict the coincidence between GWs and GRBs. To calculate the probability of detecting a GRB counterpart, we first assume the Gaussian structured jet profile model for GRB
\begin{equation}
L(\theta_{\rm V})=L_c\exp(-\frac{\theta_{\rm V}^2}{2\theta_c^2}) \,,
\label{eq:Ltheta}
\end{equation}
here $L(\theta)$ is the luminosity per unit solid angle, $\theta_{\rm V}$ is the viewing angle. $L_c$ and $\theta_c$ are the structure parameters that define the sharpness of the angular profile. The structured jet parameter is given by $\theta_c=4.7^\circ$~\cite{Howell:2018nhu}. We then assume a standard broken power law form for the distribution of the short GRB
\begin{equation}
\Phi(L)\propto
\begin{cases}
(L/L_*)^a, & L<L_* \\
(L/L_*)^b, & L\ge L_*
\end{cases}
\label{eq:PhiL}
\end{equation}
where $L$ is the isotropic rest frame luminosity in the 1-10000 keV energy range and $L_*$ is a characteristic luminosity that separates the low and high end of the luminosity function and $a$ and $b$ are the characteristic slopes describing these regimes, respectively. We set $a=-1.95$, $b=-3$ and $L_*=2\times 10^{52}~\rm erg~sec^{-1}$~\cite{Wanderman:2014eza}. Now we can sample the GW+GRB events from the probability distribution $\Phi(L)dL$. The details of the calculation can be found in~\cite{Yang:2021qge}.

Figure~\ref{fig:BNSGW} shows one realization of the GW detections and GW-GRB coincidences for 5-years observation of AEDGE by assuming a GRB detector with the characteristics of THESEUS. We can see that the number of useful bright sirens is drastically limited by the detections of GRB. Only around 32 GW-GRB cases can be obtained by THESEUS with its X-Gamma ray Imaging Spectrometer (XGIS). This result can be compared to the case of HLVKI network with Fermi-GBM satellite, for which only 14 GW-GRB cases can be detected in 10 years~\cite{Belgacem:2019tbw,Yang:2021qge}, and to the case of ET+2CE network with 300 GW-GRB cases in 5 years~\cite{Yang:2021qge}.

\begin{figure}
\centering
\includegraphics[width=0.9\textwidth]{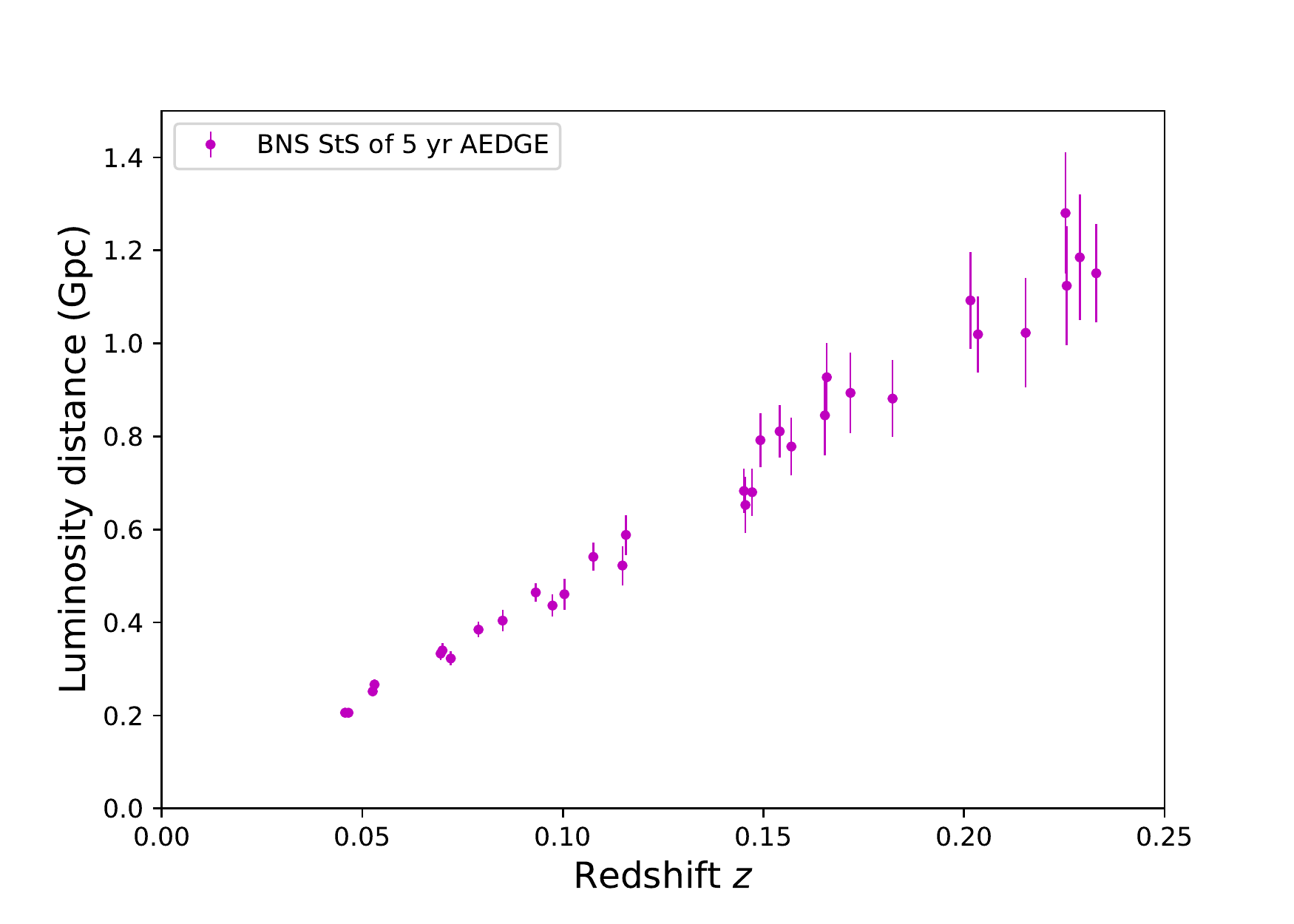}
\caption{The Hubble diagram of one realization of mock bright sirens from AEDGE. }
\label{fig:Hubble_diagram}
\end{figure}

Having the sampled GW-GRB candidates we can construct the Hubble diagram of future bright sirens by 5-years observation of AEDGE. We assume the fiducial cosmological model  to be the $\Lambda$CDM model with $H_0=67.72~\rm km~s^{-1}~Mpc^{-1}$ and $\Omega_m=0.3104$, corresponding to the mean values obtained from the latest \textit{Planck} TT,TE,EE+lowE+lensing+BAO+Pantheon data combination~\cite{Planck:2018vyg}. We also fix the present CMB temperature $T_{\rm CMB}=2.7255~\rm K$, the sum of neutrino masses $\Sigma_{\nu}m_{\nu}=0.06~\rm eV$, and the effective extra relativistic degrees of freedom $N_{\rm eff}=3.046$, as in the {\it Planck} baseline analysis.  We assign three contributions to the error of the luminosity distance, i.e., the instrumental error, the weak lensing, and the peculiar velocity of the source galaxy. For weak lensing we adopt the analytically fitting formula~\cite{Hirata:2010ba,Tamanini:2016zlh}
\begin{equation}
\left(\frac{\Delta d_L(z)}{d_L(z)}\right)_{\rm lens}=0.066\left(\frac{1-(1+z)^{-0.25}}{0.25}\right)^{1.8} \,.
\end{equation}
We consider a delensing factor, i.e., the use of dedicated matter surveys along the line of sight of the GW event in order to estimate the lensing magnification distribution and thus remove part of the uncertainty due to weak lensing, which can reduce the weak lensing uncertainty. Following~\cite{Speri:2020hwc} we adopt a phenomenological formula
\begin{equation}
F_{\rm delens}(z)=1-\frac{0.3}{\pi/2}\arctan(z/z_*) \,,
\end{equation}
where $z_*=0.073$. The final error from weak lensing is
\begin{equation}
\left(\frac{\Delta d_L(z)}{d_L(z)}\right)_{\rm delens}=F_{\rm delens}(z)\left(\frac{\Delta d_L(z)}{d_L(z)}\right)_{\rm lens} \,.
\end{equation}
For the peculiar velocity uncertainty, we use the fitting formula~\cite{Kocsis:2005vv},
\begin{equation}
\left(\frac{\Delta d_L(z)}{d_L(z)}\right)_{\rm pec}=\left[1+\frac{c(1+z)^2}{H(z)d_L(z)}\right]\frac{\sqrt{\langle v^2\rangle}}{c} \,,
\end{equation}
here we set peculiar velocity value to be 500 km/s, in agreement with average values observed in galaxy catalogs.
Finally, the instrumental error due to the parameter estimation from the matched filtering waveform is estimated as $\Delta d_L/d_L\sim$ 1/SNR~\cite{Dalal:2006qt,Li:2013lza}. In this paper we consider the BNS associated with a short GRB which is usually beamed within an angle of about $25^\circ$.  The correlation between distance and inclination is substantially broken, so the above estimate becomes accurate~\cite{Nissanke:2009kt}. The final uncertainty of the luminosity distance is just the sum of the above errors in quadrature. We construct the Hubble diagram of bright sirens from future AEDGE in Fig.~\ref{fig:Hubble_diagram}.

\section{Cosmological applications \label{sec:cosmology}}

Gravitational waves as the standard sirens have many applications in cosmology. In this paper we focus on using bright sirens to measure the Hubble constant, constraining the dynamics of dark energy, and testing the validity of GR. To measure the cosmological parameters, StS is very analogous to the SNe Ia as the standard candles. The $d_L-z$ relation of standard sirens does not rely on the calibration and the underlying physics is very clear, which are the great advantages over SNe Ia. This feature of StS makes it a perfect probe of the expansion of our Universe and hence to measure the cosmological parameters especially for Hubble constant and equation of state of dark energy.  On the other hand, StS is very sensitive to the propagation of GW when considering the modified gravity theories. So that we can test GR from the propagation of GWs across cosmological distances~\cite{Belgacem:2017ihm,Belgacem:2018lbp,Nishizawa:2017nef,Arai:2017hxj,Nishizawa:2019rra,LISACosmologyWorkingGroup:2019mwx,Belgacem:2019tbw,Belgacem:2019zzu,Mukherjee:2019wcg,DAgostino:2019hvh,Bonilla:2019mbm,Mukherjee:2020mha,Kalomenopoulos:2020klp,Mastrogiovanni:2020mvm,Mastrogiovanni:2020gua,Finke:2021aom}. In a general modified gravity theory, the linearized evolution equation for GWs traveling on an FRW background is~\cite{LISACosmologyWorkingGroup:2019mwx}
\begin{equation}
\tilde{h}_A^{\prime\prime}+2[1-\delta(\eta)]\mathcal{H}\tilde{h}_A^{\prime}+[c_T^2(\eta)k^2+m_T^2(\eta)]\tilde{h}_A=\Pi_A \,,
\label{eq:GWpropa}
\end{equation}
$\tilde{h}_A$ are the Fourier modes of the GW amplitude and $A=+,\times$ labels the two polarizations. $\mathcal{H}=a^{\prime}/a$ is the Hubble parameter in conformal time and the primes indicating derivatives with respect to conformal time $\eta$. The function $\delta (\eta)$ is introduced to modify the friction term in the propagation equation, thus denoting the effects of modified gravity theory. $c_T$ corresponds to the speed of gravitational waves.  In theories of modified gravity the tensor mode can be massive, with $m_T$ its mass. In GR we have $\delta=0$, $c_T=c$,  and $m_T=0$. The detection of GW170817/GRB170817A puts a very tight constraint $(c_T-c)/c<\mathcal{O}(10^{-15})$~\cite{LIGOScientific:2017zic}. In this paper we only consider the extra $\delta$ in the friction term. Then it is possible to show that standard sirens do not measure the usual electromagnetic luminosity distance but a gravitational-wave luminosity distance~\cite{Belgacem:2017ihm,Belgacem:2018lbp},
\begin{equation}
d_L^{\rm gw}(z)=d_L^{\rm em}(z)\exp\left\{-\int_0^z\frac{dz^{\prime}}{1+z^{\prime}}\delta(z^{\prime})\right\} \,.
\end{equation}
We can use the difference between $d_L^{\rm gw}(z)$ and $d_L^{\rm em}(z)$ as a smoking gun of the modified gravity effects.

Now we can apply the mock bright sirens by AEDGE to some cosmological applications.  As in~\cite{Yang:2021qge} we consider four parameterizations for dark energy model and/or modified gravity theory as follows:
\begin{itemize}
\item We assume the baseline $\Lambda$CDM model with a cosmological constant ($w=-1$). We would like to investigate how precisely the bright sirens of AEDGE can constrain the Hubble constant $H_0$ and the dark matter density parameter $\Omega_m$.

\item We extend the baseline $\Lambda$CDM model and assume a constant equation of state of dark energy $w$, which is denoted as $w$CDM. 

\item We consider further the dynamics of dark energy and assume a Chevallier-Polarski-Linder (CPL) form $w(z)=w_0+w_az/(1+z)$ for the equation of state~\cite{Chevallier:2000qy}.

\item We assume a phenomenological parameterization of the modified GW propagation $\Xi(z)\equiv d_L^{\rm gw}(z)/d_L^{\rm em}(z)=\Xi_0+(1-\Xi_0)/{(1+z)^n}$~\cite{Belgacem:2019tbw}, which is denoted as MG.
\end{itemize} 

\begin{figure}[tbp]
\centering
\begin{subfigure}[b]{0.45\textwidth}
\includegraphics[width=\textwidth]{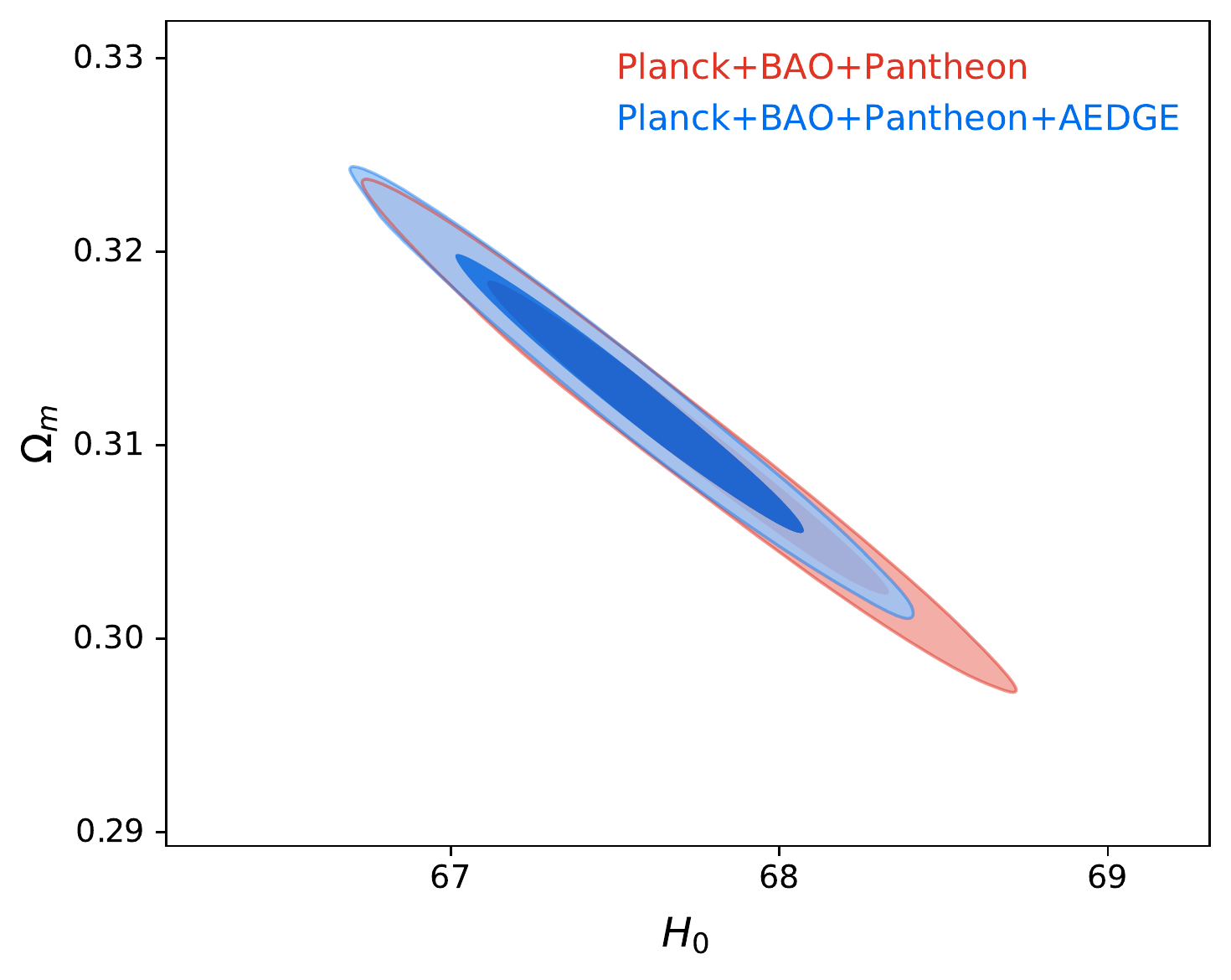}
\caption{$\Lambda$CDM}
\end{subfigure}
\hfill
\begin{subfigure}[b]{0.45\textwidth}
\includegraphics[width=\textwidth]{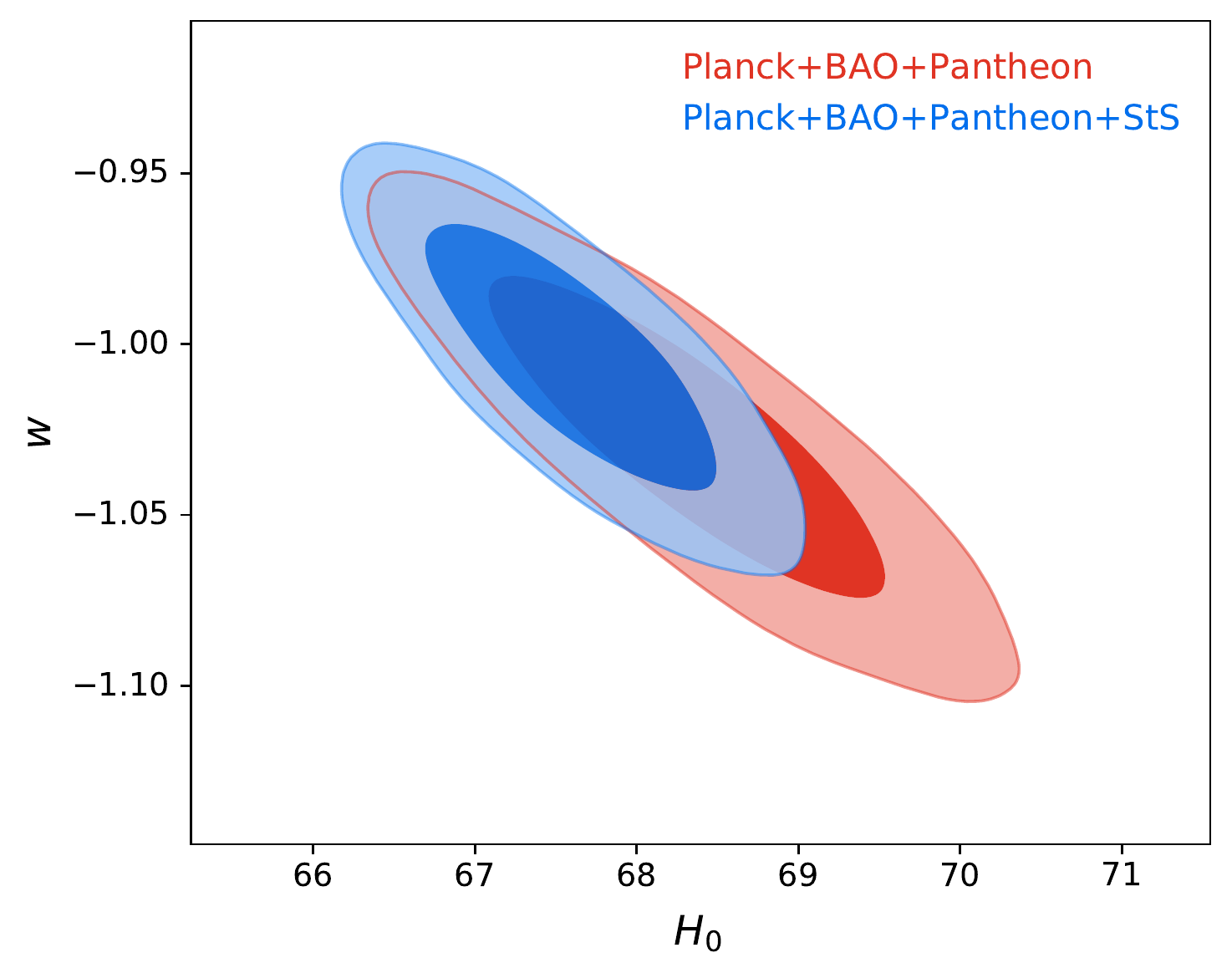}
\caption{$w$CDM}
\end{subfigure}
\\
\begin{subfigure}[b]{0.45\textwidth}
\includegraphics[width=\textwidth]{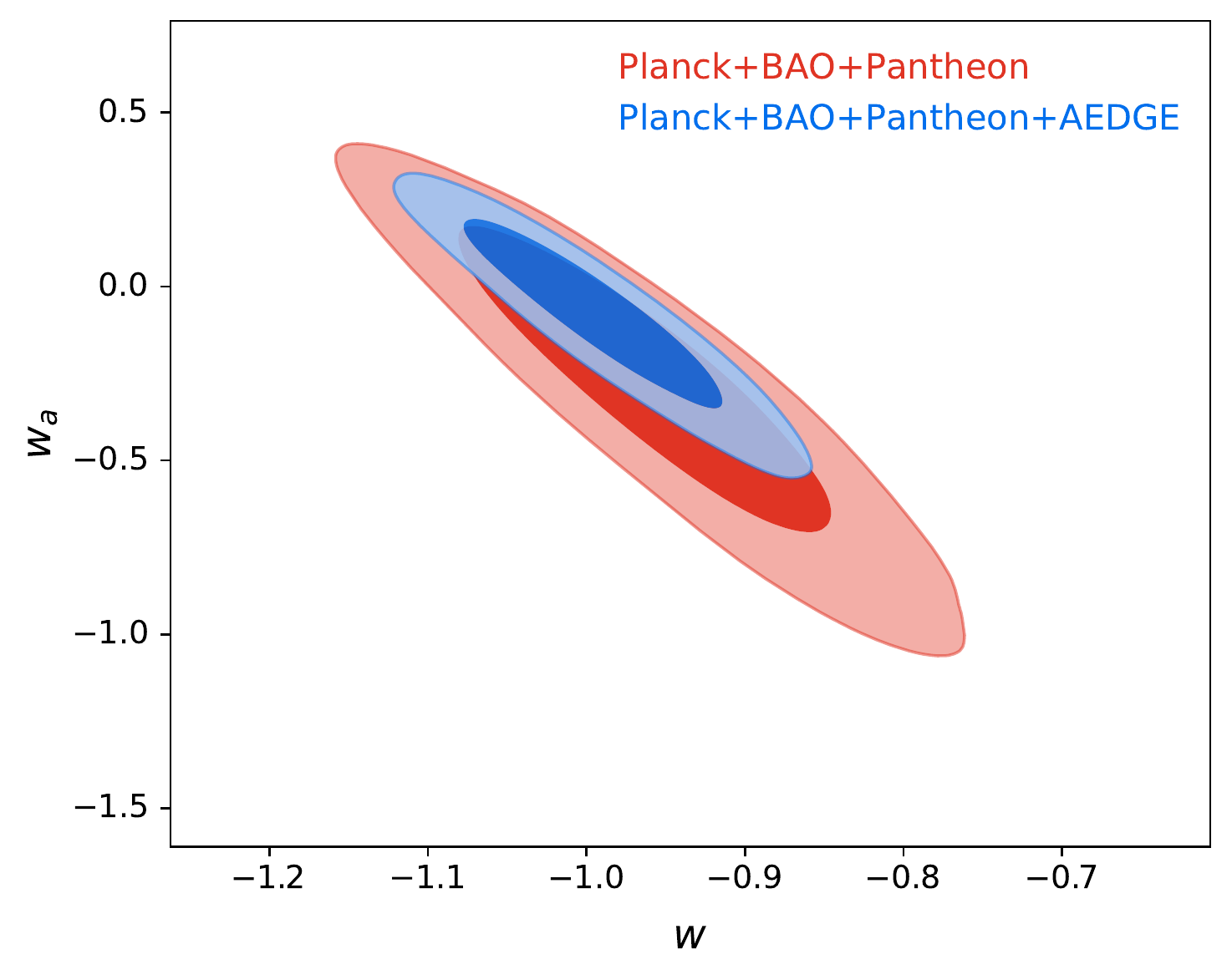}
\caption{CPL}
\end{subfigure}
\hfill
\begin{subfigure}[b]{0.45\textwidth}
\includegraphics[width=\textwidth]{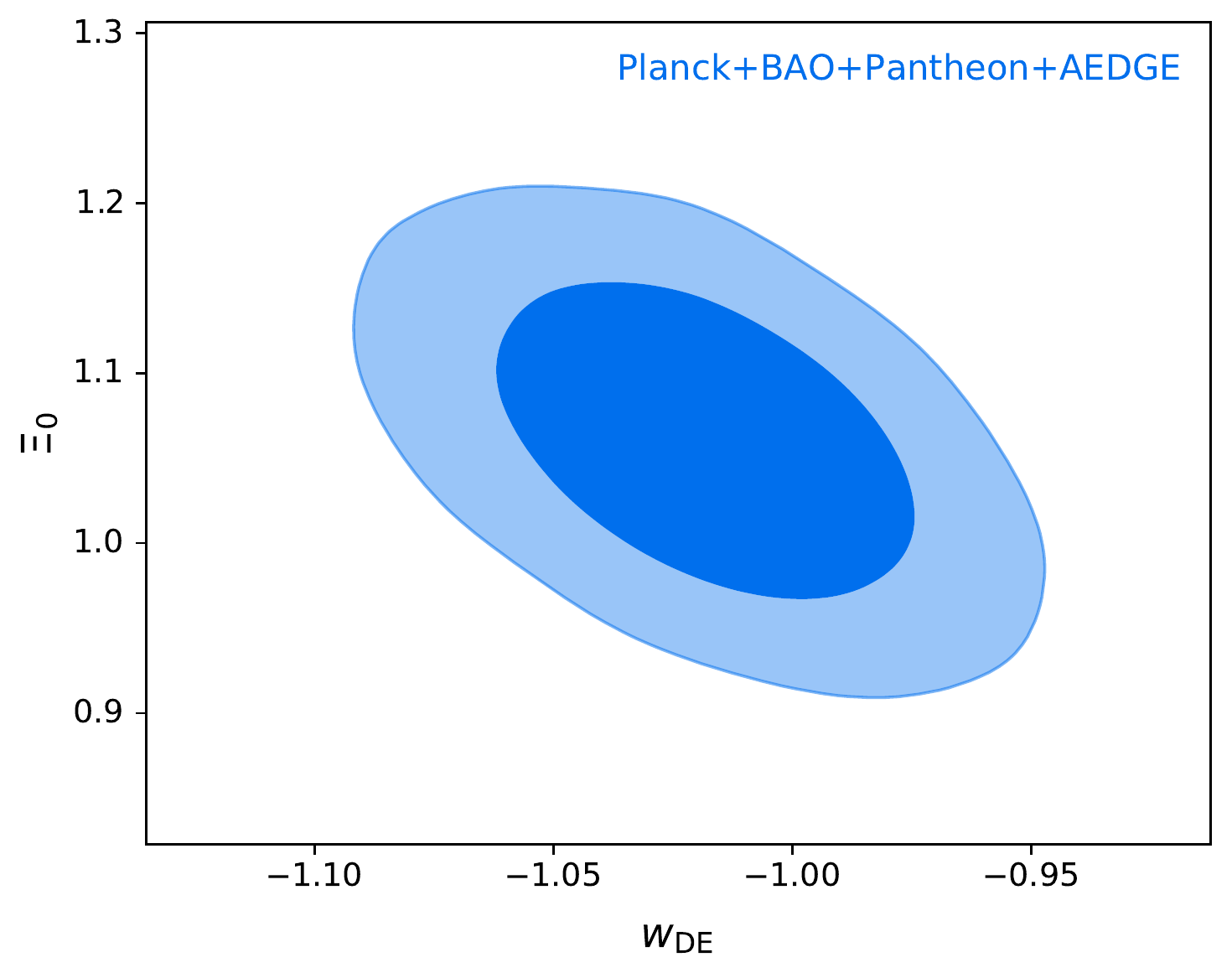}
\caption{MG}
\end{subfigure}
\caption{The constraints of the selected parameters in each DE/MG model. Contours contain 68 \% and 95 \% of the probability.}
\label{fig:MCMC}
\end{figure}

We also include the traditional EM data sets for comparison. We use CMB data from the latest {\it Planck}~\cite{Planck:2018vyg}. For BAO we adopt the isotropic constraints provided by 6dFGS at $z_{\rm eff}=0.106$~\cite{Beutler:2011hx}, SDSS-MGS DR7 at $z_{\rm eff}=0.15$~\cite{Ross:2014qpa}, and ``consensus'' BAOs in three redshift slices with effective redshifts $z_{\rm eff}$ = 0.38, 0.51, and 0.61~\cite{BOSS:2016apd,Vargas-Magana:2016imr,BOSS:2016hvq}. We use the Pantheon data~\cite{Scolnic:2017caz} as the latest compilation of SNe Ia. For each model we consider different data sets combinations, i.e., AEDGE standard sirens alone, {\it Planck}+BAO+Pantheon, and finally all together. To get the posteriors of these cosmological parameters, we run Markov-Chain Monte-Carlo (MCMC) by the package {\sc Cobaya}~\cite{Torrado:2020dgo,2019ascl.soft10019T}. The marginalized statistics of the parameters and the plots are produced by the Python package {\sc GetDist}~\cite{Lewis:2019xzd}. The contour plots of the cosmological parameters are shown in Fig.~\ref{fig:MCMC}. Note that we just show the constraints of the important parameters in each model under consideration. The detailed result for parameters is summarized in Tab.~\ref{tab:parameter}.

\begin{table}
\centering
\scalebox{0.7}{
\begin{tabular}{ll|ccccc} 
\hline\hline
\multicolumn{2}{l|}{\diagbox{Data and model}{Parameter}} & $H_0$           & $\Omega_m$                   & $w_0$                                          & $w_a$                                      & $\Xi_0$             \\ 
\hline
\multirow{3}{*}{AEDGE}                     & $\Lambda$CDM  & $66.2\pm 1.4$ & $0.46^{+0.22}_{-0.30}$           & --                                             & --                                         & --                  \\
                                         & $w$CDM        & $66.0^{+1.6}_{-2.1}$ & $0.61^{+0.35}_{-0.13}$    & $-1.5^{+1.4}_{-1.2}$ & --                                         & --                  \\
                                         & CPL   & $66.1^{+1.6}_{-2.0}$ & $0.59^{+0.33}_{-0.16}$    & $-1.6^{+1.4}_{-1.3}$                               & -- & --                  \\ 
\hline
\multirow{3}{*}{Planck+BAO+Pantheon}     & $\Lambda$CDM  & $67.72\pm 0.40$ & $0.3104\pm 0.0054$           & --                                             & --                                         & --                  \\
                                         & $w$CDM        & $68.34\pm 0.81$ & $0.3057\pm 0.0075$           & $-1.028\pm 0.031$                              & --                                         & --                  \\
                                         & CPL   & $68.31\pm 0.82$ & $0.3065\pm 0.0077$           & $-0.957\pm 0.080$                              & $-0.29^{+0.32}_{-0.26}$                    & --                  \\ 
\hline
\multirow{4}{*}{Planck+BAO+Pantheon+AEDGE} & $\Lambda$CDM  & $67.55\pm 0.35$ & $0.3126\pm 0.0047$           & --                                             & --                                         & --                  \\
                                         & $w$CDM        & $67.61\pm 0.59$ & $0.3123\pm 0.0057$           & $-1.004\pm 0.026$                              & --                                         & --                  \\
                                         & CPL   & $67.50\pm 0.61$ & $0.3138\pm 0.0061$ & $-0.944\pm 0.075$                     & $-0.23^{+0.29}_{-0.25}$                    & --                  \\
                                         & MG            & $68.07\pm 0.74$ & $0.3081\pm 0.0070$           & $-1.019\pm 0.029$                              & --                                         & $1.061\pm 0.061$  \\ 
\hline
\end{tabular}
}
\caption{The constraints of the parameters in each  DE/MG model for different data sets combinations.  The numbers are the mean values with 68\% limits of the errors.}
\label{tab:parameter}
\end{table}

\section{Conclusion and discussions \label{sec:conclusion}}

In this paper, for the first time, we consider the atom interferometer as a novel GW detector and investigate its potential to detect bright sirens and hence the applications on cosmology and modified gravity theory. We choose the space-borne AEDGE as our fiducial AI detector and assume a five-years GW data-taking period with the follow-up observation of GRB counterparts by THESEUS. Based on current knowledge of the BNS merger rates and GRB model, we estimate the number of bright sirens that AEDGE could detect in the future. We then apply these bright sirens to several cosmological applications, such as the measurement of Hubble constant, the constraint of dynamics of dark energy and the deviations from GR.

As a preliminary investigation, our results show there should be around 32 bright siren detections within 5-years observation time by AEDGE. Though about 7819 BNS cases can be observed, the joint GW-GRB events are drastically limited by the EM counterparts detectability. Note we only consider GRB as the EM counterpart in this paper. By these 32 bright sirens alone, Hubble constant can be measured with a precision of 2.1\%, which is sufficient to arbitrate the current tension between local and high-$z$ measurements of $H_0$~\cite{Chen:2017rfc}. It requires more than 50 events for a comparable level of precision by HLV (LIGO+Virgo) network~\cite{Chen:2017rfc}. Adding AEDGE to current {\it Planck}+BAO+Pantheon datasets we can improve the measurement of $H_0$ from 0.6\% to 0.5\%.  For the dynamics of dark energy, AEDGE itself can not put a tight constraint for the equation of state. This is due to the number and redshift range (only up to 0.25) of the events. However, the inclusion of AEDGE can still improve the constraints of $w$ and $w_a$, though marginally. When constraining the MG effects from the propagation of GW, AEDGE 5-years bright sirens with current most precise EM experiments can measure $\Xi_0$ at 5.7\% precision level, which is two times better than by 10-years operation of HLVKI network (however much inferior to ET+2CE network which can measure $\Xi_0$ at 0.7\% precision level)~\cite{Belgacem:2019tbw}.

In addition to BNS bright sirens, MBHB could also be possibly accompanied by an EM counterpart. They are the main targets of LISA~\cite{Klein:2015hvg,Tamanini:2016zlh}. We have also done the simulation of MBHBs observed by AEDGE. However, only the very late pre-merger stage can enter the frequency band of AEDGE. From the calculation of fisher matrix, the uncertainty of the waveform parameters including the luminosity distance and angular resolution are very large, which makes it impossible to apply the MBHBs as the standard sirens in this paper. 

We should stress here that the strategy of detecting BNS bright sirens for AEDGE is different from that of LIGO and ET which focus on the high frequency range($f>10$ Hz). AEDGE can only monitor the early stage of BNS inspiral thus not able to detect GWs and EM counterparts simultaneously. Actually the long-time observation of inspiral phase by AEDGE can constrain the GW waveform parameters very precisely (including the localization of the source) and predict the merger time accurately. It will provide not only an early warning when GWs entering the frequency band of ground-based LIs like LIGO and ET but also enough time and confidence to observe the associated EM counterpart at coalescence. In this paper we assume AEDGE could monitor the early inspiral phase of BNS in a 5-years data-taking time, then store and analyze the data properly. After a successful follow-up EM observation for one BNS and associating them correctly, we can mark them as a joint GW+EM event. We also note that since we only consider the GRB as the EM counterpart in this paper, the prediction of the number of bright sirens is very conservative in some sense. 

Though the number of BNS bright sirens is very limited, the dark sirens of BNS and BBH without EM counterparts could provide us with a huge amount of cases. From the fisher matrix calculation, we have checked that the sky localization of BNS and BBH measured by AEDGE can be order of $\mathcal{O}(10^{-1}-10^{-4})~\rm deg^2$, which is much better than the one for  LI detectors like LIGO and ET (typically around the order of $\mathcal{O}(1-1000)~\rm deg^2$). This is due to the fact that from AEDGE one can monitor BNS and BBH in the inspiral stage for a long time before coalescence. The single baseline orbits around the Earth and the Sun, which cause it to reorient and change position significantly during the lifetime of the source, and make it similar to having multiple baselines/detectors. Thus more angular information can be encoded~\cite{Graham:2017lmg}. From such precise angular resolution, the redshifts of the dark sirens by AEDGE can be estimated well by identifying the host galaxies. The specific research of dark sirens by AEDGE can be found in~\cite{Yang:2021xox}.

\acknowledgments
The authors would like to thank Ling-Feng Wang, Wen-Hong Ruan and Chang Liu for helpful discussion on the fisher matrix calculation. RGC is supported by the National Natural Science Foundation of China Grants No.11690022, No.11821505, No.11991052, No.11947302 and by the Strategic Priority Research Program  of  the  Chinese  Academy  of  Sciences  Grant  No.XDB23030100, the Key Research Program of the CAS Grant No.XDPB15, and the Key Research Program of FrontierSciences of CAS.  TY would like to thank the secondment between APCTP and ITP in May and June for the completion of this work. This work is supported by an appointment to the YST Program at the APCTP through the Science and Technology Promotion Fund and Lottery Fund of the Korean Government, and the Korean Local Governments - Gyeongsangbuk-do Province and Pohang City.

\bibliographystyle{JHEP}
\bibliography{ref}

\providecommand{\href}[2]{#2}\begingroup\raggedright\begin{thebibliography}{10}

\bibitem{LIGOScientific:2016aoc}
{\scshape LIGO Scientific, Virgo} collaboration, \emph{{Observation of
  Gravitational Waves from a Binary Black Hole Merger}},
  \href{https://doi.org/10.1103/PhysRevLett.116.061102}{\emph{Phys. Rev. Lett.}
  {\bfseries 116} (2016) 061102}
  [\href{https://arxiv.org/abs/1602.03837}{{\ttfamily 1602.03837}}].

\bibitem{LIGOScientific:2017vwq}
{\scshape LIGO Scientific, Virgo} collaboration, \emph{{GW170817: Observation
  of Gravitational Waves from a Binary Neutron Star Inspiral}},
  \href{https://doi.org/10.1103/PhysRevLett.119.161101}{\emph{Phys. Rev. Lett.}
  {\bfseries 119} (2017) 161101}
  [\href{https://arxiv.org/abs/1710.05832}{{\ttfamily 1710.05832}}].

\bibitem{LIGOScientific:2018mvr}
{\scshape LIGO Scientific, Virgo} collaboration, \emph{{GWTC-1: A
  Gravitational-Wave Transient Catalog of Compact Binary Mergers Observed by
  LIGO and Virgo during the First and Second Observing Runs}},
  \href{https://doi.org/10.1103/PhysRevX.9.031040}{\emph{Phys. Rev. X}
  {\bfseries 9} (2019) 031040}
  [\href{https://arxiv.org/abs/1811.12907}{{\ttfamily 1811.12907}}].

\bibitem{LIGOScientific:2020ibl}
{\scshape LIGO Scientific, Virgo} collaboration, \emph{{GWTC-2: Compact Binary
  Coalescences Observed by LIGO and Virgo During the First Half of the Third
  Observing Run}},
  \href{https://doi.org/10.1103/PhysRevX.11.021053}{\emph{Phys. Rev. X}
  {\bfseries 11} (2021) 021053}
  [\href{https://arxiv.org/abs/2010.14527}{{\ttfamily 2010.14527}}].

\bibitem{LIGOScientific:2021qlt}
{\scshape LIGO Scientific, KAGRA, VIRGO} collaboration, \emph{{Observation of
  Gravitational Waves from Two Neutron Star\textendash{}Black Hole
  Coalescences}},
  \href{https://doi.org/10.3847/2041-8213/ac082e}{\emph{Astrophys. J. Lett.}
  {\bfseries 915} (2021) L5}
  [\href{https://arxiv.org/abs/2106.15163}{{\ttfamily 2106.15163}}].

\bibitem{LIGOScientific:2017ync}
{\scshape LIGO Scientific, Virgo, Fermi GBM, INTEGRAL, IceCube, AstroSat
  Cadmium Zinc Telluride Imager Team, IPN, Insight-Hxmt, ANTARES, Swift, AGILE
  Team, 1M2H Team, Dark Energy Camera GW-EM, DES, DLT40, GRAWITA, Fermi-LAT,
  ATCA, ASKAP, Las Cumbres Observatory Group, OzGrav, DWF (Deeper Wider Faster
  Program), AST3, CAASTRO, VINROUGE, MASTER, J-GEM, GROWTH, JAGWAR,
  CaltechNRAO, TTU-NRAO, NuSTAR, Pan-STARRS, MAXI Team, TZAC Consortium, KU,
  Nordic Optical Telescope, ePESSTO, GROND, Texas Tech University, SALT Group,
  TOROS, BOOTES, MWA, CALET, IKI-GW Follow-up, H.E.S.S., LOFAR, LWA, HAWC,
  Pierre Auger, ALMA, Euro VLBI Team, Pi of Sky, Chandra Team at McGill
  University, DFN, ATLAS Telescopes, High Time Resolution Universe Survey,
  RIMAS, RATIR, SKA South Africa/MeerKAT} collaboration, \emph{{Multi-messenger
  Observations of a Binary Neutron Star Merger}},
  \href{https://doi.org/10.3847/2041-8213/aa91c9}{\emph{Astrophys. J. Lett.}
  {\bfseries 848} (2017) L12}
  [\href{https://arxiv.org/abs/1710.05833}{{\ttfamily 1710.05833}}].

\bibitem{LIGOScientific:2017zic}
{\scshape LIGO Scientific, Virgo, Fermi-GBM, INTEGRAL} collaboration,
  \emph{{Gravitational Waves and Gamma-rays from a Binary Neutron Star Merger:
  GW170817 and GRB 170817A}},
  \href{https://doi.org/10.3847/2041-8213/aa920c}{\emph{Astrophys. J. Lett.}
  {\bfseries 848} (2017) L13}
  [\href{https://arxiv.org/abs/1710.05834}{{\ttfamily 1710.05834}}].

\bibitem{Schutz:1999xj}
B.F.~Schutz, \emph{{Gravitational wave astronomy}},
  \href{https://doi.org/10.1088/0264-9381/16/12A/307}{\emph{Class. Quant.
  Grav.} {\bfseries 16} (1999) A131}
  [\href{https://arxiv.org/abs/gr-qc/9911034}{{\ttfamily gr-qc/9911034}}].

\bibitem{Barack:2018yly}
L.~Barack et~al., \emph{{Black holes, gravitational waves and fundamental
  physics: a roadmap}},
  \href{https://doi.org/10.1088/1361-6382/ab0587}{\emph{Class. Quant. Grav.}
  {\bfseries 36} (2019) 143001}
  [\href{https://arxiv.org/abs/1806.05195}{{\ttfamily 1806.05195}}].

\bibitem{Sasaki:2018dmp}
M.~Sasaki, T.~Suyama, T.~Tanaka and S.~Yokoyama, \emph{{Primordial black
  holes\textemdash{}perspectives in gravitational wave astronomy}},
  \href{https://doi.org/10.1088/1361-6382/aaa7b4}{\emph{Class. Quant. Grav.}
  {\bfseries 35} (2018) 063001}
  [\href{https://arxiv.org/abs/1801.05235}{{\ttfamily 1801.05235}}].

\bibitem{Gair:2012nm}
J.R.~Gair, M.~Vallisneri, S.L.~Larson and J.G.~Baker, \emph{{Testing General
  Relativity with Low-Frequency, Space-Based Gravitational-Wave Detectors}},
  \href{https://doi.org/10.12942/lrr-2013-7}{\emph{Living Rev. Rel.} {\bfseries
  16} (2013) 7} [\href{https://arxiv.org/abs/1212.5575}{{\ttfamily
  1212.5575}}].

\bibitem{Ezquiaga:2018btd}
J.M.~Ezquiaga and M.~Zumalac\'arregui, \emph{{Dark Energy in light of
  Multi-Messenger Gravitational-Wave astronomy}},
  \href{https://doi.org/10.3389/fspas.2018.00044}{\emph{Front. Astron. Space
  Sci.} {\bfseries 5} (2018) 44}
  [\href{https://arxiv.org/abs/1807.09241}{{\ttfamily 1807.09241}}].

\bibitem{Cai:2017cbj}
R.-G.~Cai, Z.~Cao, Z.-K.~Guo, S.-J.~Wang and T.~Yang, \emph{{The
  Gravitational-Wave Physics}},
  \href{https://doi.org/10.1093/nsr/nwx029}{\emph{Natl. Sci. Rev.} {\bfseries
  4} (2017) 687} [\href{https://arxiv.org/abs/1703.00187}{{\ttfamily
  1703.00187}}].

\bibitem{Meszaros:2019xej}
P.~M\'esz\'aros, D.B.~Fox, C.~Hanna and K.~Murase, \emph{{Multi-Messenger
  Astrophysics}}, \href{https://doi.org/10.1038/s42254-019-0101-z}{\emph{Nature
  Rev. Phys.} {\bfseries 1} (2019) 585}
  [\href{https://arxiv.org/abs/1906.10212}{{\ttfamily 1906.10212}}].

\bibitem{Christensen:2018iqi}
N.~Christensen, \emph{{Stochastic Gravitational Wave Backgrounds}},
  \href{https://doi.org/10.1088/1361-6633/aae6b5}{\emph{Rept. Prog. Phys.}
  {\bfseries 82} (2019) 016903}
  [\href{https://arxiv.org/abs/1811.08797}{{\ttfamily 1811.08797}}].

\bibitem{Perkins:2020tra}
S.E.~Perkins, N.~Yunes and E.~Berti, \emph{{Probing Fundamental Physics with
  Gravitational Waves: The Next Generation}},
  \href{https://doi.org/10.1103/PhysRevD.103.044024}{\emph{Phys. Rev. D}
  {\bfseries 103} (2021) 044024}
  [\href{https://arxiv.org/abs/2010.09010}{{\ttfamily 2010.09010}}].

\bibitem{Bian:2021ini}
L.~Bian et~al., \emph{{The Gravitational-Wave Physics II: Progress}},
  \href{https://doi.org/10.1007/s11433-021-1781-x}{\emph{Sci. China Phys. Mech.
  Astron.} {\bfseries 64} (2021) 120401}
  [\href{https://arxiv.org/abs/2106.10235}{{\ttfamily 2106.10235}}].

\bibitem{Hu:2017mde}
W.-R.~Hu and Y.-L.~Wu, \emph{{The Taiji Program in Space for gravitational wave
  physics and the nature of gravity}},
  \href{https://doi.org/10.1093/nsr/nwx116}{\emph{Natl. Sci. Rev.} {\bfseries
  4} (2017) 685}.

\bibitem{Ruan:2018tsw}
W.-H.~Ruan, Z.-K.~Guo, R.-G.~Cai and Y.-Z.~Zhang, \emph{{Taiji program:
  Gravitational-wave sources}},
  \href{https://doi.org/10.1142/S0217751X2050075X}{\emph{Int. J. Mod. Phys. A}
  {\bfseries 35} (2020) 2050075}
  [\href{https://arxiv.org/abs/1807.09495}{{\ttfamily 1807.09495}}].

\bibitem{TianQin:2015yph}
{\scshape TianQin} collaboration, \emph{{TianQin: a space-borne gravitational
  wave detector}},
  \href{https://doi.org/10.1088/0264-9381/33/3/035010}{\emph{Class. Quant.
  Grav.} {\bfseries 33} (2016) 035010}
  [\href{https://arxiv.org/abs/1512.02076}{{\ttfamily 1512.02076}}].

\bibitem{Dimopoulos:2007cj}
S.~Dimopoulos, P.W.~Graham, J.M.~Hogan, M.A.~Kasevich and S.~Rajendran,
  \emph{{Gravitational Wave Detection with Atom Interferometry}},
  \href{https://doi.org/10.1016/j.physletb.2009.06.011}{\emph{Phys. Lett. B}
  {\bfseries 678} (2009) 37} [\href{https://arxiv.org/abs/0712.1250}{{\ttfamily
  0712.1250}}].

\bibitem{Dimopoulos:2008sv}
S.~Dimopoulos, P.W.~Graham, J.M.~Hogan, M.A.~Kasevich and S.~Rajendran,
  \emph{{An Atomic Gravitational Wave Interferometric Sensor (AGIS)}},
  \href{https://doi.org/10.1103/PhysRevD.78.122002}{\emph{Phys. Rev. D}
  {\bfseries 78} (2008) 122002}
  [\href{https://arxiv.org/abs/0806.2125}{{\ttfamily 0806.2125}}].

\bibitem{Graham:2012sy}
P.W.~Graham, J.M.~Hogan, M.A.~Kasevich and S.~Rajendran, \emph{{A New Method
  for Gravitational Wave Detection with Atomic Sensors}},
  \href{https://doi.org/10.1103/PhysRevLett.110.171102}{\emph{Phys. Rev. Lett.}
  {\bfseries 110} (2013) 171102}
  [\href{https://arxiv.org/abs/1206.0818}{{\ttfamily 1206.0818}}].

\bibitem{Hogan:2015xla}
J.M.~Hogan and M.A.~Kasevich, \emph{{Atom interferometric gravitational wave
  detection using heterodyne laser links}},
  \href{https://doi.org/10.1103/PhysRevA.94.033632}{\emph{Phys. Rev. A}
  {\bfseries 94} (2016) 033632}
  [\href{https://arxiv.org/abs/1501.06797}{{\ttfamily 1501.06797}}].

\bibitem{Zhan:2019quq}
M.-S.~Zhan et~al., \emph{{ZAIGA: Zhaoshan Long-baseline Atom Interferometer
  Gravitation Antenna}},
  \href{https://doi.org/10.1142/S0218271819400054}{\emph{Int. J. Mod. Phys. D}
  {\bfseries 29} (2019) 1940005}
  [\href{https://arxiv.org/abs/1903.09288}{{\ttfamily 1903.09288}}].

\bibitem{Badurina:2019hst}
L.~Badurina et~al., \emph{{AION: An Atom Interferometer Observatory and
  Network}}, \href{https://doi.org/10.1088/1475-7516/2020/05/011}{\emph{JCAP}
  {\bfseries 05} (2020) 011}
  [\href{https://arxiv.org/abs/1911.11755}{{\ttfamily 1911.11755}}].

\bibitem{Geiger:2015tma}
R.~Geiger et~al., \emph{{Matter-wave laser Interferometric Gravitation Antenna
  (MIGA): New perspectives for fundamental physics and geosciences}},  in
  \emph{{50th Rencontres de Moriond on Gravitation: 100 years after GR}},
  pp.~163--172, 5, 2015 [\href{https://arxiv.org/abs/1505.07137}{{\ttfamily
  1505.07137}}].

\bibitem{Canuel:2019abg}
B.~Canuel et~al., \emph{{ELGAR\textemdash{}a European Laboratory for
  Gravitation and Atom-interferometric Research}},
  \href{https://doi.org/10.1088/1361-6382/aba80e}{\emph{Class. Quant. Grav.}
  {\bfseries 37} (2020) 225017}
  [\href{https://arxiv.org/abs/1911.03701}{{\ttfamily 1911.03701}}].

\bibitem{Graham:2017pmn}
{\scshape MAGIS} collaboration, \emph{{Mid-band gravitational wave detection
  with precision atomic sensors}},
  \href{https://arxiv.org/abs/1711.02225}{{\ttfamily 1711.02225}}.

\bibitem{AEDGE:2019nxb}
{\scshape AEDGE} collaboration, \emph{{AEDGE: Atomic Experiment for Dark Matter
  and Gravity Exploration in Space}},
  \href{https://doi.org/10.1140/epjqt/s40507-020-0080-0}{\emph{EPJ Quant.
  Technol.} {\bfseries 7} (2020) 6}
  [\href{https://arxiv.org/abs/1908.00802}{{\ttfamily 1908.00802}}].

\bibitem{Graham:2017lmg}
P.W.~Graham and S.~Jung, \emph{{Localizing Gravitational Wave Sources with
  Single-Baseline Atom Interferometers}},
  \href{https://doi.org/10.1103/PhysRevD.97.024052}{\emph{Phys. Rev. D}
  {\bfseries 97} (2018) 024052}
  [\href{https://arxiv.org/abs/1710.03269}{{\ttfamily 1710.03269}}].

\bibitem{Ellis:2020lxl}
J.~Ellis and V.~Vaskonen, \emph{{Probes of gravitational waves with atom
  interferometers}},
  \href{https://doi.org/10.1103/PhysRevD.101.124013}{\emph{Phys. Rev. D}
  {\bfseries 101} (2020) 124013}
  [\href{https://arxiv.org/abs/2003.13480}{{\ttfamily 2003.13480}}].

\bibitem{Dalal:2006qt}
N.~Dalal, D.E.~Holz, S.A.~Hughes and B.~Jain, \emph{{Short grb and binary black
  hole standard sirens as a probe of dark energy}},
  \href{https://doi.org/10.1103/PhysRevD.74.063006}{\emph{Phys. Rev. D}
  {\bfseries 74} (2006) 063006}
  [\href{https://arxiv.org/abs/astro-ph/0601275}{{\ttfamily
  astro-ph/0601275}}].

\bibitem{Nissanke:2009kt}
S.~Nissanke, D.E.~Holz, S.A.~Hughes, N.~Dalal and J.L.~Sievers,
  \emph{{Exploring short gamma-ray bursts as gravitational-wave standard
  sirens}}, \href{https://doi.org/10.1088/0004-637X/725/1/496}{\emph{Astrophys.
  J.} {\bfseries 725} (2010) 496}
  [\href{https://arxiv.org/abs/0904.1017}{{\ttfamily 0904.1017}}].

\bibitem{Zhao:2010sz}
W.~Zhao, C.~Van Den~Broeck, D.~Baskaran and T.G.F.~Li, \emph{{Determination of
  Dark Energy by the Einstein Telescope: Comparing with CMB, BAO and SNIa
  Observations}}, \href{https://doi.org/10.1103/PhysRevD.83.023005}{\emph{Phys.
  Rev. D} {\bfseries 83} (2011) 023005}
  [\href{https://arxiv.org/abs/1009.0206}{{\ttfamily 1009.0206}}].

\bibitem{Cai:2016sby}
R.-G.~Cai and T.~Yang, \emph{{Estimating cosmological parameters by the
  simulated data of gravitational waves from the Einstein Telescope}},
  \href{https://doi.org/10.1103/PhysRevD.95.044024}{\emph{Phys. Rev. D}
  {\bfseries 95} (2017) 044024}
  [\href{https://arxiv.org/abs/1608.08008}{{\ttfamily 1608.08008}}].

\bibitem{Cai:2017yww}
R.-G.~Cai, N.~Tamanini and T.~Yang, \emph{{Reconstructing the dark sector
  interaction with LISA}},
  \href{https://doi.org/10.1088/1475-7516/2017/05/031}{\emph{JCAP} {\bfseries
  05} (2017) 031} [\href{https://arxiv.org/abs/1703.07323}{{\ttfamily
  1703.07323}}].

\bibitem{Yang:2021qge}
T.~Yang, \emph{{Gravitational-Wave Detector Networks: Standard Sirens on
  Cosmology and Modified Gravity Theory}},
  \href{https://doi.org/10.1088/1475-7516/2021/05/044}{\emph{JCAP} {\bfseries
  05} (2021) 044} [\href{https://arxiv.org/abs/2103.01923}{{\ttfamily
  2103.01923}}].

\bibitem{Vitale:2018yhm}
S.~Vitale, W.M.~Farr, K.~Ng and C.L.~Rodriguez, \emph{{Measuring the star
  formation rate with gravitational waves from binary black holes}},
  \href{https://doi.org/10.3847/2041-8213/ab50c0}{\emph{Astrophys. J. Lett.}
  {\bfseries 886} (2019) L1}
  [\href{https://arxiv.org/abs/1808.00901}{{\ttfamily 1808.00901}}].

\bibitem{Belgacem:2019tbw}
E.~Belgacem, Y.~Dirian, S.~Foffa, E.J.~Howell, M.~Maggiore and T.~Regimbau,
  \emph{{Cosmology and dark energy from joint gravitational wave-GRB
  observations}},
  \href{https://doi.org/10.1088/1475-7516/2019/08/015}{\emph{JCAP} {\bfseries
  08} (2019) 015} [\href{https://arxiv.org/abs/1907.01487}{{\ttfamily
  1907.01487}}].

\bibitem{Madau:2014bja}
P.~Madau and M.~Dickinson, \emph{{Cosmic Star Formation History}},
  \href{https://doi.org/10.1146/annurev-astro-081811-125615}{\emph{Ann. Rev.
  Astron. Astrophys.} {\bfseries 52} (2014) 415}
  [\href{https://arxiv.org/abs/1403.0007}{{\ttfamily 1403.0007}}].

\bibitem{Cutler:1994ys}
C.~Cutler and E.E.~Flanagan, \emph{{Gravitational waves from merging compact
  binaries: How accurately can one extract the binary's parameters from the
  inspiral wave form?}},
  \href{https://doi.org/10.1103/PhysRevD.49.2658}{\emph{Phys. Rev. D}
  {\bfseries 49} (1994) 2658}
  [\href{https://arxiv.org/abs/gr-qc/9402014}{{\ttfamily gr-qc/9402014}}].

\bibitem{THESEUS:2017qvx}
{\scshape THESEUS} collaboration, \emph{{The THESEUS space mission concept:
  science case, design and expected performances}},
  \href{https://doi.org/10.1016/j.asr.2018.03.010}{\emph{Adv. Space Res.}
  {\bfseries 62} (2018) 191}
  [\href{https://arxiv.org/abs/1710.04638}{{\ttfamily 1710.04638}}].

\bibitem{THESEUS:2017wvz}
{\scshape THESEUS} collaboration, \emph{{THESEUS: a key space mission concept
  for Multi-Messenger Astrophysics}},
  \href{https://doi.org/10.1016/j.asr.2018.04.013}{\emph{Adv. Space Res.}
  {\bfseries 62} (2018) 662}
  [\href{https://arxiv.org/abs/1712.08153}{{\ttfamily 1712.08153}}].

\bibitem{Howell:2018nhu}
E.J.~Howell, K.~Ackley, A.~Rowlinson and D.~Coward, \emph{{Joint gravitational
  wave -- gamma-ray burst detection rates in the aftermath of GW170817}},
  \href{https://arxiv.org/abs/1811.09168}{{\ttfamily 1811.09168}}.

\bibitem{Wanderman:2014eza}
D.~Wanderman and T.~Piran, \emph{{The rate, luminosity function and time delay
  of non-Collapsar short GRBs}},
  \href{https://doi.org/10.1093/mnras/stv123}{\emph{Mon. Not. Roy. Astron.
  Soc.} {\bfseries 448} (2015) 3026}
  [\href{https://arxiv.org/abs/1405.5878}{{\ttfamily 1405.5878}}].

\bibitem{Planck:2018vyg}
{\scshape Planck} collaboration, \emph{{Planck 2018 results. VI. Cosmological
  parameters}},
  \href{https://doi.org/10.1051/0004-6361/201833910}{\emph{Astron. Astrophys.}
  {\bfseries 641} (2020) A6}
  [\href{https://arxiv.org/abs/1807.06209}{{\ttfamily 1807.06209}}].

\bibitem{Hirata:2010ba}
C.M.~Hirata, D.E.~Holz and C.~Cutler, \emph{{Reducing the weak lensing noise
  for the gravitational wave Hubble diagram using the non-Gaussianity of the
  magnification distribution}},
  \href{https://doi.org/10.1103/PhysRevD.81.124046}{\emph{Phys. Rev. D}
  {\bfseries 81} (2010) 124046}
  [\href{https://arxiv.org/abs/1004.3988}{{\ttfamily 1004.3988}}].

\bibitem{Tamanini:2016zlh}
N.~Tamanini, C.~Caprini, E.~Barausse, A.~Sesana, A.~Klein and A.~Petiteau,
  \emph{{Science with the space-based interferometer eLISA. III: Probing the
  expansion of the Universe using gravitational wave standard sirens}},
  \href{https://doi.org/10.1088/1475-7516/2016/04/002}{\emph{JCAP} {\bfseries
  04} (2016) 002} [\href{https://arxiv.org/abs/1601.07112}{{\ttfamily
  1601.07112}}].

\bibitem{Speri:2020hwc}
L.~Speri, N.~Tamanini, R.R.~Caldwell, J.R.~Gair and B.~Wang, \emph{{Testing the
  Quasar Hubble Diagram with LISA Standard Sirens}},
  \href{https://doi.org/10.1103/PhysRevD.103.083526}{\emph{Phys. Rev. D}
  {\bfseries 103} (2021) 083526}
  [\href{https://arxiv.org/abs/2010.09049}{{\ttfamily 2010.09049}}].

\bibitem{Kocsis:2005vv}
B.~Kocsis, Z.~Frei, Z.~Haiman and K.~Menou, \emph{{Finding the electromagnetic
  counterparts of cosmological standard sirens}},
  \href{https://doi.org/10.1086/498236}{\emph{Astrophys. J.} {\bfseries 637}
  (2006) 27} [\href{https://arxiv.org/abs/astro-ph/0505394}{{\ttfamily
  astro-ph/0505394}}].

\bibitem{Li:2013lza}
T.G.F.~Li, \emph{{Extracting Physics from Gravitational Waves: Testing the
  Strong-field Dynamics of General Relativity and Inferring the Large-scale
  Structure of the Universe}}, Ph.D. thesis, Vrije U., Amsterdam, 2013.

\bibitem{Belgacem:2017ihm}
E.~Belgacem, Y.~Dirian, S.~Foffa and M.~Maggiore, \emph{{Gravitational-wave
  luminosity distance in modified gravity theories}},
  \href{https://doi.org/10.1103/PhysRevD.97.104066}{\emph{Phys. Rev. D}
  {\bfseries 97} (2018) 104066}
  [\href{https://arxiv.org/abs/1712.08108}{{\ttfamily 1712.08108}}].

\bibitem{Belgacem:2018lbp}
E.~Belgacem, Y.~Dirian, S.~Foffa and M.~Maggiore, \emph{{Modified
  gravitational-wave propagation and standard sirens}},
  \href{https://doi.org/10.1103/PhysRevD.98.023510}{\emph{Phys. Rev. D}
  {\bfseries 98} (2018) 023510}
  [\href{https://arxiv.org/abs/1805.08731}{{\ttfamily 1805.08731}}].

\bibitem{Nishizawa:2017nef}
A.~Nishizawa, \emph{{Generalized framework for testing gravity with
  gravitational-wave propagation. I. Formulation}},
  \href{https://doi.org/10.1103/PhysRevD.97.104037}{\emph{Phys. Rev. D}
  {\bfseries 97} (2018) 104037}
  [\href{https://arxiv.org/abs/1710.04825}{{\ttfamily 1710.04825}}].

\bibitem{Arai:2017hxj}
S.~Arai and A.~Nishizawa, \emph{{Generalized framework for testing gravity with
  gravitational-wave propagation. II. Constraints on Horndeski theory}},
  \href{https://doi.org/10.1103/PhysRevD.97.104038}{\emph{Phys. Rev. D}
  {\bfseries 97} (2018) 104038}
  [\href{https://arxiv.org/abs/1711.03776}{{\ttfamily 1711.03776}}].

\bibitem{Nishizawa:2019rra}
A.~Nishizawa and S.~Arai, \emph{{Generalized framework for testing gravity with
  gravitational-wave propagation. III. Future prospect}},
  \href{https://doi.org/10.1103/PhysRevD.99.104038}{\emph{Phys. Rev. D}
  {\bfseries 99} (2019) 104038}
  [\href{https://arxiv.org/abs/1901.08249}{{\ttfamily 1901.08249}}].

\bibitem{LISACosmologyWorkingGroup:2019mwx}
{\scshape LISA Cosmology Working Group} collaboration, \emph{{Testing modified
  gravity at cosmological distances with LISA standard sirens}},
  \href{https://doi.org/10.1088/1475-7516/2019/07/024}{\emph{JCAP} {\bfseries
  07} (2019) 024} [\href{https://arxiv.org/abs/1906.01593}{{\ttfamily
  1906.01593}}].

\bibitem{Belgacem:2019zzu}
E.~Belgacem, S.~Foffa, M.~Maggiore and T.~Yang, \emph{{Gaussian processes
  reconstruction of modified gravitational wave propagation}},
  \href{https://doi.org/10.1103/PhysRevD.101.063505}{\emph{Phys. Rev. D}
  {\bfseries 101} (2020) 063505}
  [\href{https://arxiv.org/abs/1911.11497}{{\ttfamily 1911.11497}}].

\bibitem{Mukherjee:2019wcg}
S.~Mukherjee, B.D.~Wandelt and J.~Silk, \emph{{Probing the theory of gravity
  with gravitational lensing of gravitational waves and galaxy surveys}},
  \href{https://doi.org/10.1093/mnras/staa827}{\emph{Mon. Not. Roy. Astron.
  Soc.} {\bfseries 494} (2020) 1956}
  [\href{https://arxiv.org/abs/1908.08951}{{\ttfamily 1908.08951}}].

\bibitem{DAgostino:2019hvh}
R.~D'Agostino and R.C.~Nunes, \emph{{Probing observational bounds on
  scalar-tensor theories from standard sirens}},
  \href{https://doi.org/10.1103/PhysRevD.100.044041}{\emph{Phys. Rev. D}
  {\bfseries 100} (2019) 044041}
  [\href{https://arxiv.org/abs/1907.05516}{{\ttfamily 1907.05516}}].

\bibitem{Bonilla:2019mbm}
A.~Bonilla, R.~D'Agostino, R.C.~Nunes and J.C.N.~de~Araujo, \emph{{Forecasts on
  the speed of gravitational waves at high $z$}},
  \href{https://doi.org/10.1088/1475-7516/2020/03/015}{\emph{JCAP} {\bfseries
  03} (2020) 015} [\href{https://arxiv.org/abs/1910.05631}{{\ttfamily
  1910.05631}}].

\bibitem{Mukherjee:2020mha}
S.~Mukherjee, B.D.~Wandelt and J.~Silk, \emph{{Testing the general theory of
  relativity using gravitational wave propagation from dark standard sirens}},
  \href{https://doi.org/10.1093/mnras/stab001}{\emph{Mon. Not. Roy. Astron.
  Soc.} {\bfseries 502} (2021) 1136}
  [\href{https://arxiv.org/abs/2012.15316}{{\ttfamily 2012.15316}}].

\bibitem{Kalomenopoulos:2020klp}
M.~Kalomenopoulos, S.~Khochfar, J.~Gair and S.~Arai, \emph{{Mapping the
  inhomogeneous Universe with standard sirens: degeneracy between inhomogeneity
  and modified gravity theories}},
  \href{https://doi.org/10.1093/mnras/stab557}{\emph{Mon. Not. Roy. Astron.
  Soc.} {\bfseries 503} (2021) 3179}
  [\href{https://arxiv.org/abs/2007.15020}{{\ttfamily 2007.15020}}].

\bibitem{Mastrogiovanni:2020mvm}
S.~Mastrogiovanni, L.~Haegel, C.~Karathanasis, I.M.n.~Hernandez and D.A.~Steer,
  \emph{{Gravitational wave friction in light of GW170817 and GW190521}},
  \href{https://doi.org/10.1088/1475-7516/2021/02/043}{\emph{JCAP} {\bfseries
  02} (2021) 043} [\href{https://arxiv.org/abs/2010.04047}{{\ttfamily
  2010.04047}}].

\bibitem{Mastrogiovanni:2020gua}
S.~Mastrogiovanni, D.~Steer and M.~Barsuglia, \emph{{Probing modified gravity
  theories and cosmology using gravitational-waves and associated
  electromagnetic counterparts}},
  \href{https://doi.org/10.1103/PhysRevD.102.044009}{\emph{Phys. Rev. D}
  {\bfseries 102} (2020) 044009}
  [\href{https://arxiv.org/abs/2004.01632}{{\ttfamily 2004.01632}}].

\bibitem{Finke:2021aom}
A.~Finke, S.~Foffa, F.~Iacovelli, M.~Maggiore and M.~Mancarella,
  \emph{{Cosmology with LIGO/Virgo dark sirens: Hubble parameter and modified
  gravitational wave propagation}},
  \href{https://arxiv.org/abs/2101.12660}{{\ttfamily 2101.12660}}.

\bibitem{Chevallier:2000qy}
M.~Chevallier and D.~Polarski, \emph{{Accelerating universes with scaling dark
  matter}}, \href{https://doi.org/10.1142/S0218271801000822}{\emph{Int. J. Mod.
  Phys. D} {\bfseries 10} (2001) 213}
  [\href{https://arxiv.org/abs/gr-qc/0009008}{{\ttfamily gr-qc/0009008}}].

\bibitem{Beutler:2011hx}
F.~Beutler, C.~Blake, M.~Colless, D.H.~Jones, L.~Staveley-Smith, L.~Campbell
  et~al., \emph{{The 6dF Galaxy Survey: Baryon Acoustic Oscillations and the
  Local Hubble Constant}},
  \href{https://doi.org/10.1111/j.1365-2966.2011.19250.x}{\emph{Mon. Not. Roy.
  Astron. Soc.} {\bfseries 416} (2011) 3017}
  [\href{https://arxiv.org/abs/1106.3366}{{\ttfamily 1106.3366}}].

\bibitem{Ross:2014qpa}
A.J.~Ross, L.~Samushia, C.~Howlett, W.J.~Percival, A.~Burden and M.~Manera,
  \emph{{The clustering of the SDSS DR7 main Galaxy sample \textendash{} I. A 4
  per cent distance measure at $z = 0.15$}},
  \href{https://doi.org/10.1093/mnras/stv154}{\emph{Mon. Not. Roy. Astron.
  Soc.} {\bfseries 449} (2015) 835}
  [\href{https://arxiv.org/abs/1409.3242}{{\ttfamily 1409.3242}}].

\bibitem{BOSS:2016apd}
{\scshape BOSS} collaboration, \emph{{The clustering of galaxies in the
  completed SDSS-III Baryon Oscillation Spectroscopic Survey: Observational
  systematics and baryon acoustic oscillations in the correlation function}},
  \href{https://doi.org/10.1093/mnras/stw2372}{\emph{Mon. Not. Roy. Astron.
  Soc.} {\bfseries 464} (2017) 1168}
  [\href{https://arxiv.org/abs/1607.03145}{{\ttfamily 1607.03145}}].

\bibitem{Vargas-Magana:2016imr}
M.~Vargas-Maga\~na et~al., \emph{{The clustering of galaxies in the completed
  SDSS-III Baryon Oscillation Spectroscopic Survey: theoretical systematics and
  Baryon Acoustic Oscillations in the galaxy correlation function}},
  \href{https://doi.org/10.1093/mnras/sty571}{\emph{Mon. Not. Roy. Astron.
  Soc.} {\bfseries 477} (2018) 1153}
  [\href{https://arxiv.org/abs/1610.03506}{{\ttfamily 1610.03506}}].

\bibitem{BOSS:2016hvq}
{\scshape BOSS} collaboration, \emph{{The clustering of galaxies in the
  completed SDSS-III Baryon Oscillation Spectroscopic Survey: baryon acoustic
  oscillations in the Fourier space}},
  \href{https://doi.org/10.1093/mnras/stw2373}{\emph{Mon. Not. Roy. Astron.
  Soc.} {\bfseries 464} (2017) 3409}
  [\href{https://arxiv.org/abs/1607.03149}{{\ttfamily 1607.03149}}].

\bibitem{Scolnic:2017caz}
D.M.~Scolnic et~al., \emph{{The Complete Light-curve Sample of
  Spectroscopically Confirmed SNe Ia from Pan-STARRS1 and Cosmological
  Constraints from the Combined Pantheon Sample}},
  \href{https://doi.org/10.3847/1538-4357/aab9bb}{\emph{Astrophys. J.}
  {\bfseries 859} (2018) 101}
  [\href{https://arxiv.org/abs/1710.00845}{{\ttfamily 1710.00845}}].

\bibitem{Torrado:2020dgo}
J.~Torrado and A.~Lewis, \emph{{Cobaya: Code for Bayesian Analysis of
  hierarchical physical models}},
  \href{https://doi.org/10.1088/1475-7516/2021/05/057}{\emph{JCAP} {\bfseries
  05} (2021) 057} [\href{https://arxiv.org/abs/2005.05290}{{\ttfamily
  2005.05290}}].

\bibitem{2019ascl.soft10019T}
J.~{Torrado} and A.~{Lewis}, \emph{{Cobaya: Bayesian analysis in cosmology}},
  Oct., 2019.

\bibitem{Lewis:2019xzd}
A.~Lewis, \emph{{GetDist: a Python package for analysing Monte Carlo samples}},
   \href{https://arxiv.org/abs/1910.13970}{{\ttfamily 1910.13970}}.

\bibitem{Chen:2017rfc}
H.-Y.~Chen, M.~Fishbach and D.E.~Holz, \emph{{A two per cent Hubble constant
  measurement from standard sirens within five years}},
  \href{https://doi.org/10.1038/s41586-018-0606-0}{\emph{Nature} {\bfseries
  562} (2018) 545} [\href{https://arxiv.org/abs/1712.06531}{{\ttfamily
  1712.06531}}].

\bibitem{Klein:2015hvg}
A.~Klein et~al., \emph{{Science with the space-based interferometer eLISA:
  Supermassive black hole binaries}},
  \href{https://doi.org/10.1103/PhysRevD.93.024003}{\emph{Phys. Rev. D}
  {\bfseries 93} (2016) 024003}
  [\href{https://arxiv.org/abs/1511.05581}{{\ttfamily 1511.05581}}].

\bibitem{Yang:2021xox}
T.~Yang, H.M.~Lee, R.-G.~Cai, H.-g.~Choi and S.~Jung, \emph{{Space-borne Atom
  Interferometric Gravitational Wave Detections II: Dark Sirens and Finding the
  One}},  \href{https://arxiv.org/abs/2110.09967}{{\ttfamily 2110.09967}}.

\end{thebibliography}\endgroup

\end{document}